\documentclass [11pt] {article}
\usepackage{authblk}

\usepackage{epsfig}

\usepackage{amssymb}

\newtheorem{thm}{Theorem}
\newtheorem{corollary}[thm]{Corollary}
\newtheorem{lemma}[thm]{Lemma}
\newtheorem{definition}{Definition}

\begin{document}



\title{Partitioning RNAs into pseudonotted and pseudoknot-free regions modeled as Dual Graphs}

\author{Louis Petingi\thanks{College of Staten Island, City University of New York, Computer Science Department, Staten Island, New York, email: louis.petingi@csi.cuny.edu. Corresponding Author} { and}  Tamar Schlick\thanks{New York University, Department of Chemistry and Courant Institute of Mathematical Sciences, New York, New York, email: sclick@nyu.edu}}
\maketitle
\begin{abstract} Dual graphs have been applied
 to model RNA secondary structures. The purpose of the paper is two-fold: we present  new graph-theoretic properties of dual graphs to validate the further analysis and classification of RNAs using these topological representations;  we also present a linear-time algorithm to partition dual graphs into topological components called {\it blocks} and determine if each block contains a {\it pseudoknot} or not. We show that a block contains a pseudoknot if and only if the block has a vertex of degree $3$ or more; this characterization allows us to efficiently isolate smaller RNA fragments and classify them as pseudoknotted or pseudoknot-free regions, while keeping these sub-structures intact. Even though non-topological techniques to detect and classify  pseudoknots have been efficiently applied, structural properties of dual graphs provide a unique perspective for the further analysis of RNAs.  Applications to RNA design can be envisioned since modular building blocks with intact pseudoknots can be combined to form new constructs.\end{abstract}
{\bf Keywords:} Graph Theory,  RNA Secondary Structures,  Partitioning,  Bi-connectivity, Pseudoknots.

\section{Introduction}
\label{S0} 

Graph theory is a well-established field of mathematics with applications to areas where the objects can be modeled as discrete structures called {\it graphs} or networks. The study of combinatorial properties of these networks, such as communication, chemical, and biological networks, can be guided by graph-theoretical principles and algorithms. Specific examples include the study of chemical structures (e.g., hydrocarbons, drug compounds)~\cite{Johnson,Mandado}, 
genetic and cellular relationships~\cite{Gunsalus,Milo}, and transportation networks~\cite{Santi14}.

In mathematical terms, an undirected graph $G=(V,E)$ is a discrete object described by a finite set of {\it vertices} $V$ and a set $E$ of unordered pair of vertices called {\it edges}, where each edge represents a connection between two vertices. 

The graphs  described in this paper were introduced in 2003 by  Gan et. al ~\cite{gan03}, called dual graphs, were applied to model RNA secondary structures (2D). The 2D elements of RNA molecules consist of double-stranded (stem) 
 regions by base pairing such as Adenine-Uracil, Guanine-Cytosine, 
 Guanine-Uracil, and single stranded loops;  stems  and loops are mapped to the vertices and edges of the corresponding dual graph, respectively. Dual graphs are needed to represent pseudoknots, structures involving an interwining of two-base-paired regions of the RNA. These are common elements in many biologically important RNAs.

Given a graph $G=(V,E)$, let the degree of a vertex $u \in V$ be the number of edges incident at $u$ in $G$. In this paper we introduce a partitioning algorithm for dual graph representations of  RNA 2D structures to recognize pseudoknots.  Our algorithm partitions a dual graph into graph-theoretic components called {\it blocks} and then determines whether each block contains a pseudoknot; we show that a block contains a pseudoknot if and only if the block has a vertex of degree $3$ or more. Thus our methodology provides a systematic approach to partition an RNA 2D structure, modeled as a dual graph, into smaller RNA regions containing pseudoknots, while providing a new topological perspective for the analysis of RNAs.

Pseudoknots can be classified into two main groups: {\it standard} and {\it recursive} pseudoknots~\cite{Dost2008,wong}. The latter is distinguished from the former by having nested pseudoknots within a  pseudoknot. While our partitioning algorithm can detect general pseudoknots, it is not within the scope of this work to classify them. With extensions of our graph-theoretical techniques, however, it may be possible to analyze and treat standard and the more  complex recursive pseudoknots structures further, as needed for specific biological applications.

In the next section, we present background material relevant to this paper, as well as notation and mathematical definitions of RNA primary, secondary, and of pseudoknot structures. In Section~\ref{S3} we describe our partitioning approach of a dual graph $G$ into components $G' \subseteq G$ called blocks, as well as new combinatorial properties of dual graphs. In Section~\ref{S4}, we characterize these blocks, and show that a block contains a pseudoknot in the RNA 2D structure if and only if the block has a vertex of degree $3$ or more. This characterization permits us to isolate pseudoknots, without {\it breaking} them so their structural properties can be further studied; moreover we also present an alternative way to visualize the presence of pseudoknots in blocks by considering their planar {\it geometric-duals}~(\cite{harary}, pg.113). In Section~\ref{S5} we illustrate algorithmic tests performed on dual graph representations of existing RNA motifs, and classify each block of a dual graph as having a pseudoknot or not.  We summarize the findings and outline new directions in Section~\ref{S6}. An Appendix section includes definitions, mathematical proofs, and supporting material.

\section{RNA background and definitions}
\label{S1} \vspace{-4pt}
Modeling of RNAs as graphs began in 1978 when Waterman introduced topological representations of RNA to analyze the secondary structure of tRNA~\cite{waterman}. In 1990, Shapiro and coworkers used a tree representation of 2D structures to measure structural similarities~\cite{shapiro}.

\begin{figure}[t]
\begin{center}
\includegraphics [scale=0.5]{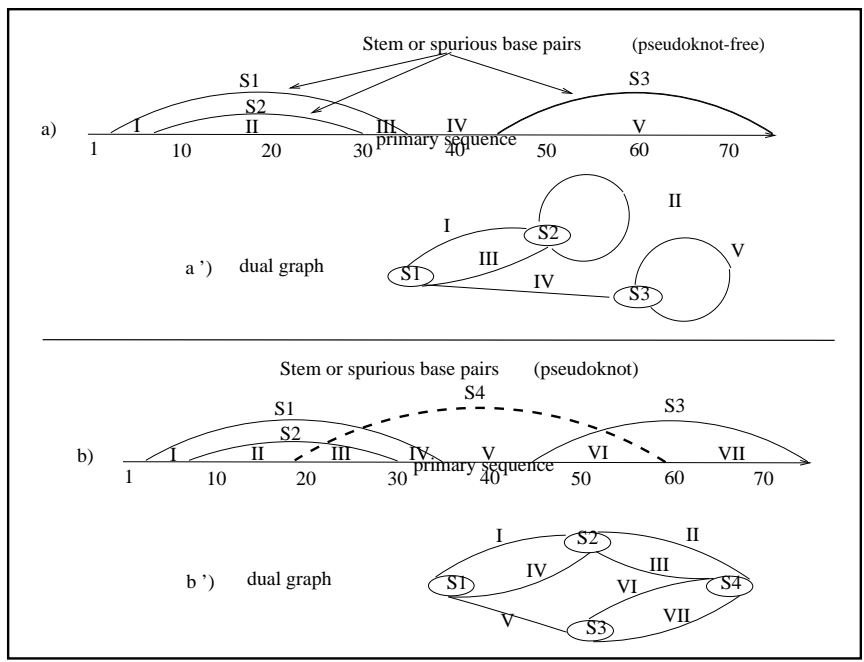}
\end{center}
\caption{ {\it  Graphical  and dual graph representations of an RNA 2D structure.}  \hspace {20.0 mm}
(a) graphical representation of a pseudoknot-free RNA primary sequence and embedded stems or base pairs; (a$^\prime$) corresponding dual graph representation.  (b) graphical representation of a pseudoknotted RNA 2D structure; (b$^\prime$) corresponding dual graph.} \label{fig:eulerian}
\end{figure}

 In 2003, Gan et. al introduced {\it tree} and {\it dual} graph-theoretic representations of RNA 2D motifs in a framework called RAG (RNA-As-Graphs)~\cite{fera04,gan03,gan04,izzo11}. While tree graphs are intuitive and easily applied to many areas of RNA research such as partitioning (e.g.,  Kim et al.~\cite{Kim14}),  complex RNA secondary structures with pseudoknots ({\it PKs}) can only be represented by dual graphs (see \cite{Kim13-1} for a survey on these topological representations); a pseudoknot is an intertwining of two-based-paired regions (stems) of an RNA (see Figure~\ref{fig:eulerian}).

The structural configuration of pseudoknots does not lend itself well to computational detection due to its overlapping nature. The base pairing in pseudoknots is not well-nested making the presence of pseudoknots in RNA sequences more difficult to predict by the dynamic programming~\cite{Dirks04} and context-free grammars standard methods \cite{Condon04}. Our methodology, based on topological properties of dual graphs, provides a new perspective for the  problem of detection and classification of pseudoknots and of general RNAs.  

Following (Kravchenko, 2009~\cite{Kra09}), we define our biological variables as follows.

\begin {definition} \label{def1} General terms: 

\begin {enumerate}
\item [i.] {\it RNA primary structure}: a sequence of linearly ordered bases $ x_1, x_2, \ldots, x_r$, where $x_i \in \{ A, U,$ $ C, G\}$.
\item [ii.] {\it canonical base pair}: a base pair $(x_i, x_j) \in \{(A,U), (U,A), (C,G), (G,C),$ 
                 $(G,U), (U,G)\}$.
\item [iii.] {\it RNA secondary structure without pseudoknot - or regular structure, encapsulated in the region $(i_0, \dots, k_0)$}: an RNA 2D structure  in which no two base pairs $(x_i, x_j), (x_l, x_m)$, satisfy $i_0 \leq i < l < j < m \leq m_0$ (i.e., no two base pairs intertwined).
\item [iv.] {\it a base pair stem}: a tuple $(x_i, x_{i+1},\ldots,x_{i+r}, x_{i+(r+1)},\ldots, x_{j-1}, x_j)$ in which $(x_i, x_j),$ $ (x_{i+1}, x_{j-1}), \ldots, (x_{i+r}, x_{i+(r+1)})$ form base pairs.
\item [v.]  {\it loop region}: a tuple  $(x_1, x_2, \ldots, x_r)$ in which $\forall_{i \leq j \leq r}(x_i, x_j)$ {\it does not} form a base pair.
\item [vi.] {\it a pseudoknot encapsulated in the region} $(i_0, \dots, k_0)$:  if $\exists l,m,  (i_0 < l < m < k_0)$ such that $ (x_{i_0}, x_m)$ and $ (x_l, x_{k_0})$ are base pairs.
\end {enumerate}
\end{definition}

A graphical representation is an intuitive and natural way to depict an RNA 2D structure (see Figure~\ref{fig:eulerian}-(a),(b)), in which the $x$-axis is labeled according to the primary linearly ordered sequence of bases (Definition~\ref{def1}-i), and a stem (Definition~\ref{def1}-iv) is represented by arcs connecting base pairs. A region on the $x$-axis between the end-points of the arcs representing stems is called a {\it segment}.
 
A dual graph can be equivalently defined from the graphical representation of an RNA 2D structure as follows (Figure~\ref{fig:eulerian}). 

\begin {definition}\label{def3-a} 
The  dual graph is derived  by mapping stems and the segments between stems  ($x$-axis), of the graphical representation of an RNA 2D structure, to the vertices and edges of the dual graph, respectively. 
\end {definition}

In the next section we propose our partitioning approach of a dual graph $G$, into subgraphs $G' \subseteq G$, called blocks. In addition we show new combinatorial properties of dual graphs that will be later invoked to prove that a block contains a pseudoknot if and only if the block has a vertex of degree $3$ or more. 
 
\section{Partitioning of a graph into blocks and combinatorial properties of dual graphs}
\label{S3} \vspace{-4pt}

Let $G=(V,E)$ be a connected graph;  the following definitions will be used below. Unless otherwise stated, we follow the notation of Harary~\cite{harary},

\begin {definition} \label{def5}  Connectivity

\begin{itemize}
\item[i.] A vertex-set $X \subseteq V$ is a vertex-disconnecting set if deletion of $X$ from $G$, denoted by $G-X$, results in a disconnected graph.
\item [ii.] A vertex $v$ is an articulation point or cut-vertex if $G-{v}$ results in a disconnected graph (i.e., at least two components remain).
\item [iii.] The vertex-connectivity, $\kappa (G)$, is the minimum number of vertices whose removal from $G$ results in a disconnected graph or in a isolated vertex. If $G$ is a single edge, then $\kappa (G)=1$.
\item [iv.] A connected component  is non-separable if it does not have an articulation point (or cut-vertex). Please note that single edges or isolated points are non-separable.
\item [v.] A block is a maximal (edge-wise) non-separable graph.

\end{itemize}
\end{definition}

\begin{figure}[bth]
\begin{center}
\includegraphics [scale=0.59]{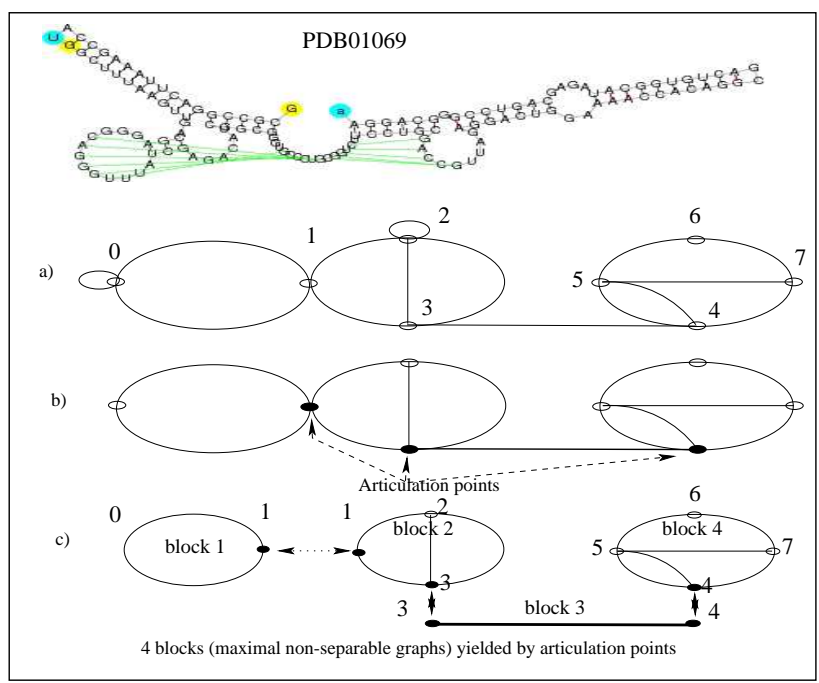}
\end{center}
\caption{\it Identification of articulation points and partitioning of the dual graph corresponding to PDB01069 RNA 2D (Catalytic Ribozyme RNA) into blocks.} \label{fig:PDB01069}
\end{figure}

The concepts of articulation points and maximal non-separable components (blocks) are related.  Indeed, articulation points partition any graph into blocks (see Fig.~\ref{fig:PDB01069}). Except when a graph $G$ is composed of just two vertices adjacent by one or several (parallel) edges ($\kappa (G) =1$), any maximal non-separable graph is bi-connected (i.e.,  $\kappa (G) \ge 2$). The fact that blocks are maximally non-separable subgraphs allows us to isolate pseudoknots, without breaking their structural properties.


Our partitioning algorithm is based on the classical result for identifying block components in a connected undirected graph introduced to John Hopcroft and Robert Tarjan (1973, \cite {HT1}), that runs in linear computational time.
 
We next state new combinatorial properties associated with dual graphs that will be invoked to show that a block contains a pseudoknot if and only if it  has a vertex of degree at least $3$. 

A graph is {\it Eulerian} if there exist a trail (see Definition~\ref{def8}-iii, Appendix~\ref{AP0}) from a vertex $v_0$ of $G$, ending at vertex $v_k$, covering all the edges of the topology; if $v_0 = v_k$ then the graph is an {\it Eulerian cycle} (see~\cite{harary}, pg. 64). Dual graph representations of general RNA 2D structures, and specifically of PKs, can be easily shown to be Eulerian graphs from Definition~\ref{def3-a}. By starting from the origin on the $x$-axis of the graphical representation and traversing to the right, a unique trail in its dual graph can be described,  where all edges are covered.
\begin{lemma}\label{eulerian1} The dual graph representations of  RNA 2D structures and of PKs are Eulerian.
\end{lemma}
As depicted in Figure~\ref{fig:eulerian}-(b), the alternating sequence of stems and segments $\{S_1, I, S_2, II, S_4, III, S_2, IV, S_1, V, S_3, VI, $ $S_4, VII,$   $ S_3\}$ of the graphical representation (b) forms an Eulerian trail in its dual graph (b$^\prime$). 


As a consequence of Lemma~\ref{eulerian1}, it follows that at most two vertices are of odd degree in a dual graph representation of an RNA 2D structure or a PK (see \cite{harary}, pg. 64). From this point on, we delete self-loops in dual graphs as they correspond to stems, in a graphical representation, not containing or not being crossed (intertwined)  by another stem. In addition under the assumption that two different stems cannot share the same bases, each endpoint of a stem in the graphical representation can be adjacent to at most two other stems, and thus the maximum degree of a vertex of a dual graph is $4$.

These facts together with Lemma~\ref{eulerian1}, yield the following corollary that will be later referenced to prove our main results.

\begin{corollary}\label{properties} Dual graph representation of an RNA 2D structure (PKs)
has the following properties:
\begin{enumerate}
\item The graph is Eulerian.
\item The maximum degree of any vertex is four.
\item The graph has at most two vertices of odd degree.
\end{enumerate}
\end{corollary}

\section{Mapping PKs to blocks  with certain degrees and combinatorial properties of dual graphs}
\label{S4} 
\subsection {Partitioning of dual graphs into pseudonotted and pseudoknot-free blocks}
\label{S4.1} 
Now we can prove that once a dual graph representing an RNA 2D structure has been partitioned into blocks, a block is contains a pseudoknot if and only if it contains a vertex of degree $3$ or more. The mathematical proofs of the lemmas stated in this section, are shown in Appendix~\ref{AP1}.

In preparation to the main results of this chapter, we first define the following.
\begin {definition} \label{blockd}
For any graph $G$, blocks can be partitioned into three classes,
\begin {enumerate}
\item Single edges.
\item Cycles.
\item Blocks containing a vertex $v$ of degree at least $3$.
\end{enumerate}
\end{definition}

From Definition~\ref{def1}-iii, an RNA 2D structure is regular (pseudoknot-free) and encapsulated in a region $(i_0, \dots, k_0)$, if no two base pairs $(x_i, x_j), (x_l, x_m)$, satisfy $i < l < j < m$, $i_0 \leq i,j,l,m \leq m_0$. Under the previous assumption that self-loops are deleted, this definition yields the following lemma,
\begin{lemma}\label{regular} Each block in the dual graph representation of a regular RNA 2D structure is either a bridge or a cycle of length $ l, l \ge 2$ (see Definition~\ref{blockd}-1,2).
\end{lemma}

Conversely we show the following.

\begin{lemma}\label{regular-pk} If an RNA 2D structure contains a pseudoknot, then its corresponding dual graph contains a block having a vertex of degree $3$ or more (see Definition~\ref{blockd}-3).
\end{lemma}

Lemma~\ref{regular} and Lemma~\ref{regular-pk} yield our main result as follows.
\begin{corollary}\label{conclude} Given a dual graph representation of  RNA 2D structure, a block represents a pseudoknot  if and only if the block has a vertex of degree $3$ or more.
\end{corollary}

To summarize our partitioning algorithm, we performed the following steps.
\begin {itemize}
\item [1.] {\it Partition the dual graph into blocks by application of Hopcroft and Tarjan's algorithm, as described in Section~\ref{S3}.}
\item [2.] {\it Analyze each block to determine if it has a vertex of degree at least 3. If that is the case then the block contains a pseudoknot, according to Corollary~\ref{conclude}.
         If not then the block represents a pseudoknot-free structure.}
\end{itemize}
\subsection {Other relevant combinatorial properties of dual graphs}

The analysis required to identify and study different types of pseudoknots, based on combinatorial properties of dual graphs, seems to be justified. For example every dual graph known so far has a planar embedding, i.e., it can be drawn in a plane so no two edges cross, even though to show planarity for general dual graphs remains an open problem.  The planarity property of dual graphs is not only supported by simple observations but also by the fact that the degree of any vertex of a dual graph is at most four. Every planar graph $G$ can be divided into regions, where each edge in $E$ divides exactly two regions of $G$. The {\it geometric-dual}\footnote{geometric-duals of planar graphs are also known as "dual-graphs" in the literature, which unfortunately coincide with the denomination of the graph representations of RNA 2D discussed in this paper.}  (see~\cite{harary}, pg. 113) representation $G_d= (V_d,E_d)$ of a planar graph $G$ is obtained as follows: for each region of the graph $G$ we have a corresponding vertex in $G_d$, and if two regions of the graph $G$ have a common edge bordering them, then the corresponding vertices in $G_d$ will be adjacent.

\begin{figure}[bth]
\begin{center}
\includegraphics [scale=0.50]{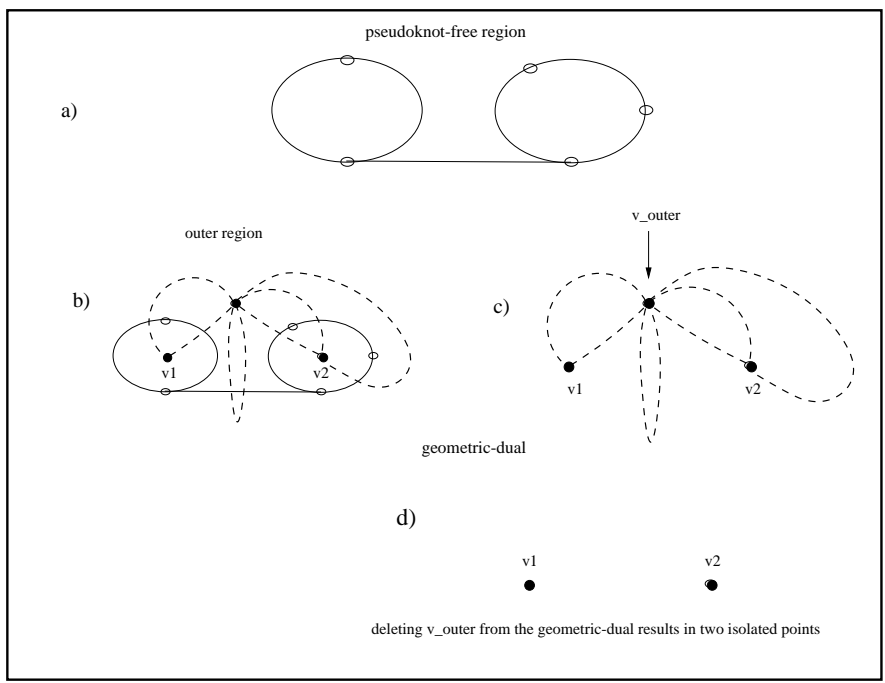}
\end{center}
\caption{{\it A pseudoknot-free region}. a) dual graph representation of a pseudoknot-free RNA; b) and c) corresponding geometric-dual; d) deleting $v_{outer}$ results in just isolated points.} \label{fig:geo1}
\end{figure}

\begin{figure}[bth]
\begin{center}
\includegraphics [scale=0.50]{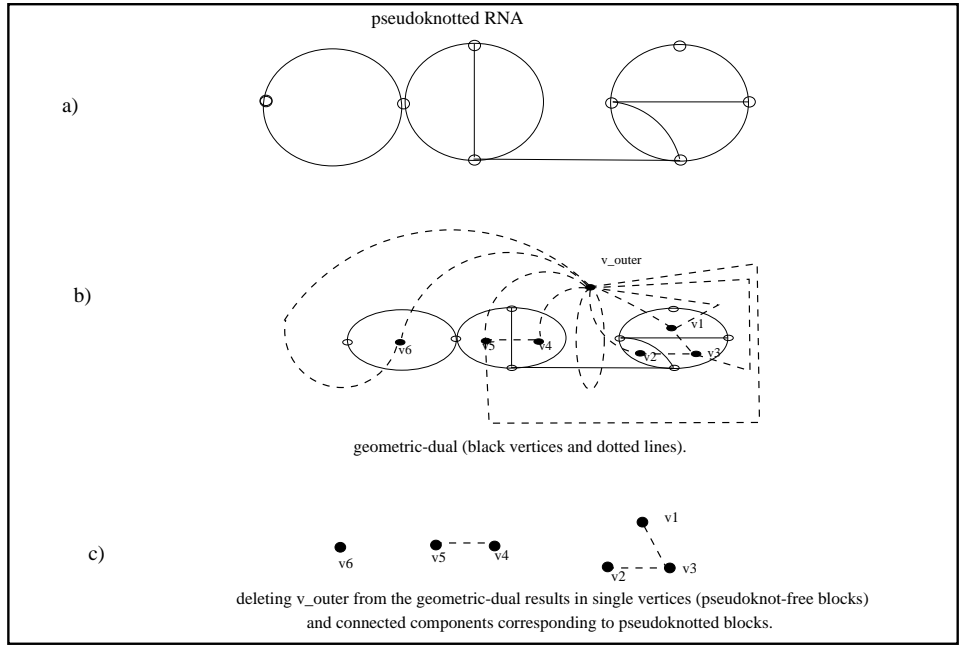}
\end{center}
\caption{{\it A pseudoknotted region}. a) dual graph representation of a pseudoknotted  RNA; b)  corresponding geometric-dual $G_d$; c) deleting $v_{outer}$ from $G_d$ results in either isolated vertices corresponding to pseudoknot-free blocks or connected components (pseudoknotted blocks).} \label{fig:geo2}
\end{figure}

It can be easily shown, from Corollary~\ref{conclude}, that the geometric-dual of a graph $G$ representing a pseudoknot-free region (under the assumption that $G$ is planar) is composed of a vertex $v_{outer}$, corresponding to the outer-region of $G$,  and $v_{outer}$ is adjacent to a vertex $v$ by $r$ parallel edges, if $v$ maps to a cycle (phase) on $r$ vertices in $G$ (bridges in $G$ correspond to self-loops in $G_d$ emanating from $v_{outer}$). Consequently deleting the vertex $v_{outer}$ from $G_d$ results in just isolated vertices (see Fig.~\ref{fig:geo1}).
Conversely, deleting $v_{outer}$ from $G_d$ when $G$ represents a pseudoknotted region, results in a graph containing connected components with at least two vertices corresponding to pseudoknotted blocks (see Fig.~\ref{fig:geo2}). These facts strongly suggest that the use of dual graphs and their geometric-duals, can shed light into the connectivity and degree analyses of standard pseudoknots (see Definitions~\ref{def3} and~\ref{def4}, Appendix~\ref{AP2}), as well as more complex pseudoknotted structures (i.e., recursive pseudoknots); it also provides a fertile theoretical ground for the study of RNA structure and function.

\section{Experimental results}
\label{S5}
\begin{figure}[bth]
\begin{center}
\includegraphics [scale=0.55]{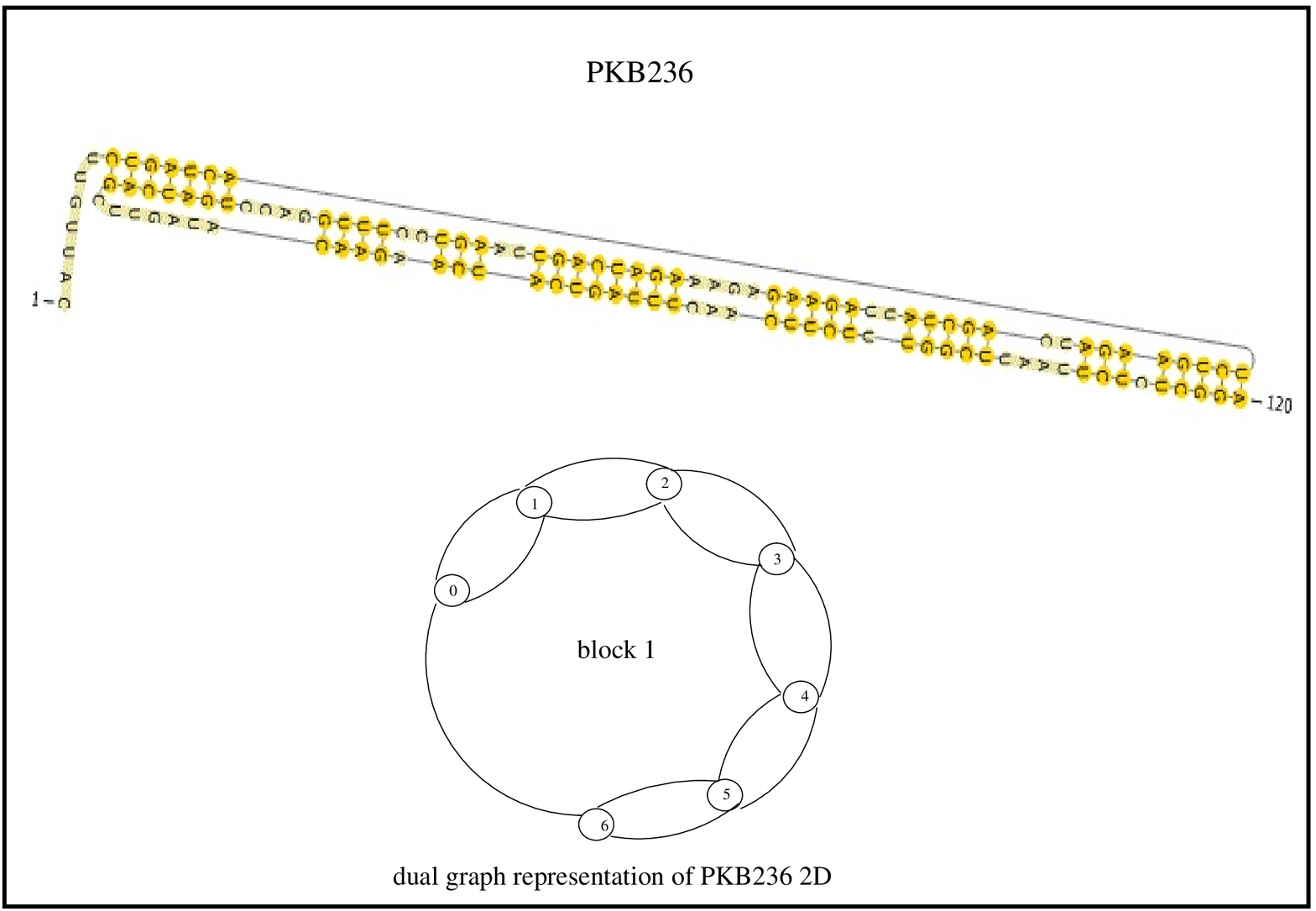}
\end{center}
\caption{{\it Partition of the dual graph corresponding to motif PKB236 (Regulatory Pseudoknot of the Interferon-gamma gene 5$^\prime$-UTR).}} \label{fig:PKB236}
\end{figure}
We illustrate our partitioning algorithm, described  at the end of Section~\ref{S4.1}, on the dual graph representations of two RNA 2D structures, based on the New York University's
  RAG database \cite{izzo11}. Our partitioning algorithm was implemented in C++ and run on a Hewlett-Packard Pavilion Dv6 (2.4 GHz)  notebook. Each partitioning
 takes less than one second because of the linear computational complexity of Hopcroft and Tarjan's algorithm.

Consider the PDB01069 RNA 2D structure, {\it Post-Cleavage State of the Thermoanaerobacter Tengcongenis GlmS Ribozyme}, known to be the only catalytic RNA to require a small-molecule activator for catalysis (see Klein et al.~\cite{Klein06}). Its dual graph is decomposed into 4 blocks as illustrated in Figure~\ref{fig:PDB01069}. According to Corollary~\ref{conclude}, block 1 and block 3, a cycle and an edge, respectively,  correspond to regular regions, while blocks 2 and 4, correspond to pseudoknots. \\
We next consider the dual graph representation of PKB236 (see Fig.~\ref{fig:PKB236}), {\it Regulatory Pseudoknot of the Interferon-gamma Gene 5$^\prime$-UTR}, thought to be involved in regulatory translation (see Ben-Asouli et al.~\cite{Ben02}); in this case the only block is the dual graph itself. As this block contains a vertex of degree $3$ or more, then this block is a pseudoknot.

Appendix~\ref{AP3} depicts the output generated when our algorithm was run on the aforementioned RNA structures.

\section {Concluding remarks and future work}\label{S6}
We have  presented a partitioning approach of the dual graph representation of RNA 2D structures into maximal non-separable components called blocks. Partitioning of a graph into blocks can be efficiently accomplished by application of
Hopcroft and Tarjan's algorithm to identify articulation points. From mathematical definitions of RNA 2D structures and of pseudoknots, we proved that an RNA 2D structure contains a pseudoknot if and only if the dual graph representation has a block in which one of the vertices is of degree $3$ or more. 
Our partitioning algorithm suggests that dual graphs along with the graph-theoretic properties described in the paper can help analyze connectivity and features of both standard (see Appendix~\ref{AP2}) and recursive pseudoknots.
Ultimately partitioning and classification of dual graphs could guide the discovery of modular regions of RNA and thus be exploited for design of novel RNAs constructed from these building blocks.

\vspace{10pt} \noindent
{\bf Acknowledgements:} The work of Tamar Schlick is supported by the National Institutes of General Medical Sciences, National Institutes of Health, awards GM100469 and GM081410. Louis Petingi's collaboration is partially supported by PSC-CUNY Grant \# 68318-00 46 from the City University of New York Research Foundation. An earlier version of this work was deposited  as a technical report~\cite {petingi15}.

\section {Appendix}

\subsection {Graph Theory definitions}
\label{AP0}

Let $G=(V,E)$ be a graph with vertex-set $V$ and edge-set $E$. We next present general graph-theoretic definitions following Harary~\cite{harary}.

\begin {definition} \label{def8}  General graph-theoretic terms:
\begin{enumerate}
\item [i.] Let $H_1.x.H_2$ represent the graph composed of two graphs, $H_1$, and $H_2$, sharing the same vertex $x$.
\item [ii.] A walk between two vertices $u$ and $v$ in graph $G=(V,E)$, is an alternating sequence of vertices and edges $<v_o=u, e_1, v_1,\ldots, e_k, v_k=v>$ such that $e_i =( v_{i-1},v_i)$ is an edge of $G$.
\item [iii.] A trail between two vertices $u$ and $v$ in graph $G=(V,E)$, is a walk between $u$ and $v$ with no repetition of edges.
\item [iv.] A path between two vertices $u$ and $v$ in graph $G=(V,E)$, is a walk (or trail) between $u$ and $v$ with no repetition of vertices.
\item [v.] A graph is {\it Eulerian} if there exist a trail from a vertex $v_0$ of $G$, ending at vertex $v_k$, covering all the edges of the topology, and if $v_0 = v_k$ then the graph is an Eulerian cycle.
\end {enumerate}
\end {definition}

\subsection {Proofs of lemmas stated in Section~\ref{S4.1}}
\label{AP1}
\setcounter{thm}{2}
\begin{figure}[bth]
\begin{center}
\includegraphics [scale=0.47]{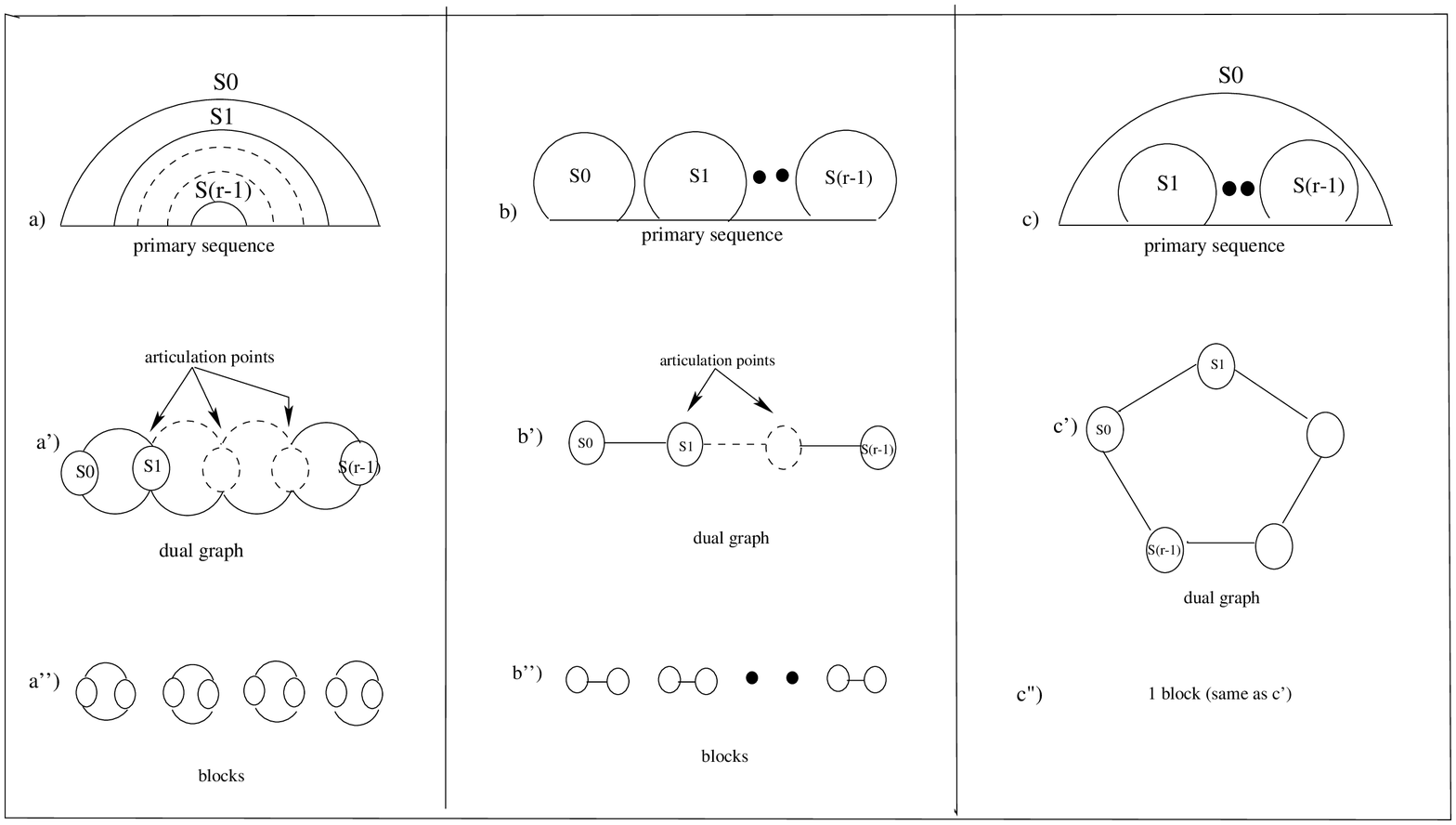}
\end{center}
\caption{ {\it Classification of PK-free regions and graphical/dual/block  representations.} (a) graphical, dual, and block representations of  $r$  nested-stems  - (b)  graphical, dual, and block representations of $r$  adjacent stems - (c) graphical, dual, and block representations of a stem containing $r-1$  adjacent stems.} \label{fig:regular1}
\end{figure}

From Definition~\ref{def1}-iii, an RNA 2D structure is regular (pseudoknot-free) and encapsulated in a region $(i_0, \dots, k_0)$, if no two base pairs $(x_i, x_j), (x_l, x_m)$, satisfy $i < l < j < m$, $i_0 \leq i,j,l,m \leq m_0$. Under the previous assumption that self-loops are deleted, this definition yields the following lemma,

\begin{lemma}\label{regular} Each block in the dual graph representation of a regular RNA 2D structure is either a bridge or a cycle of length $ l, l \ge 2$ (see Definition~\ref{blockd}-1,2).
\end{lemma}
{\bf Proof.} Consider the  graphical representation of a regular RNA 2D structure; we will proceed by construction. A regular (pseudoknot-free) region can be recursively defined as follows (see Fig.~\ref{fig:regular1}): (a) a region composed of $r$ nested-stems; (b) $r$ adjacent stems, (c) a stem containing a sequence of $r-1$ adjacent stems;  (d) a single stem (represented as an isolated vertex in its dual graph, not illustrated in  Fig.~\ref{fig:regular1}). In a transformation, a set of stems identified by properties $a$, $ b$, and $c$ in the graphical representation,  are reduced (converted) into a single stem, while its corresponding dual graph is generated (see Definition~\ref{def3-a}).
The blocks obtained from the dual graph representations of these properties, are either cycles of length $2$, single edges, a cycle of length $r$, or an isolated vertex, respectively. Consider a sequence of transformations of dual graphs  $G_1 \Rightarrow G_2 \Rightarrow \ldots \Rightarrow G_n$, where the dual graph $G_{i + 1}$ is obtained from dual graph $G_{i}$ by following the precedence rules in which, first, internal stems of the ones identified by properties (a) through (c) of the graphical representation are reduced into a single stem, while the corresponding dual graph is generated; in the dual graph we distinguish the vertex corresponding to the outer-stem. Because only distinguished vertices could be later made adjacent to other vertices in a transformation, the blocks generated by the sequence of transformations from $G_1$ through $G_{n-1}$ will remain blocks in $G_n$, with the possible addition of blocks composed of single edges. $\blacksquare$

\begin{figure}[bth]
\begin{center}
\includegraphics [scale=0.55]{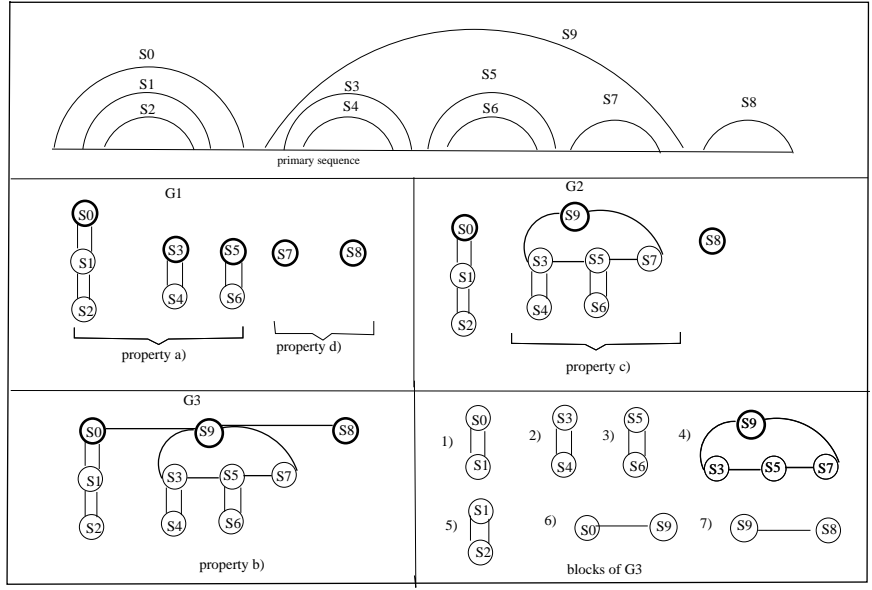}
\end{center}
\caption{\it An example illustrating Lemma~\ref{regular}.} \label{fig:regular2}
\end{figure}

To illustrate Lemma~\ref{regular}, consider Figure~\ref{fig:regular2} depicting the graphical representation of a pseudoknot-free region. The stems $S_0, S_1,$ and $S_2$, identified by property (a), with corresponding dual graph with distinguished vertex $S_0$,  are then reduced into a single stem in the graphical representation. Similarly, by property (a), we reduce the pairs of nested-stems $S_3, S_4$, and $S_5, S_6$, to two single stems with distinguished vertices $S_3$ and $S_5$, in the dual graph, respectively.  As the stem $S_9$ contains a sequence of $3$ (reduced) stems, by application of property (c), it can be reduced to a single stem with dual graph composed of a cycle on 4 vertices, and distinguished vertex $S_9$. Finally, by property (b), we connect the sequence of $3$ (reduced) stems (i.e., $S_0$, $S_9$, and $S_8$) by single edges in the dual graph.

Conversely we show the following.

\begin{figure}[bth]
\begin{center}
\includegraphics [scale=0.55]{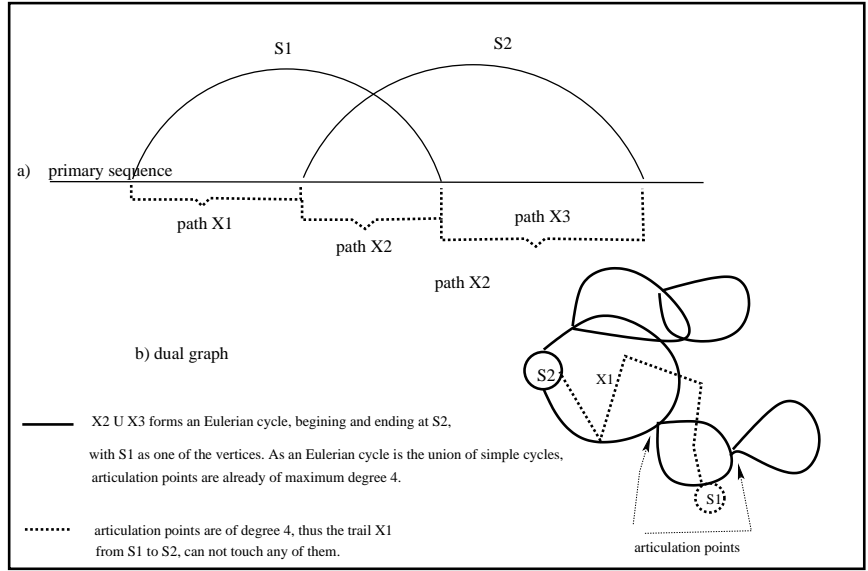}
\end{center}
\caption{\it Supporting illustration for the proof of Lemma~\ref{regular-pk}} \label{fig:regular-pk-1}
\end{figure}

\begin{lemma}\label{regular-pk} If an RNA 2D structure contains a pseudoknot, then its corresponding dual graph contains a block having a vertex of degree $3$ or more (see Definition~\ref{blockd}-3).
\end{lemma}
{\bf Proof.} By Definition~\ref{def1}-vi, if an RNA 2D structure contains a pseudoknot, there exist a stem {\it crossing} (interweaving) another stem.  Let us denominate these interweaving stems,  in the graphical representation, $S_1$ and $S_2$, respectively. There exist then three independent paths, $X_1$, $X_2$, and $X_3$, from $S_1$ to $S_2$  (see Figure~\ref{fig:regular-pk-1}-(a)), following the primary sequence of the graphical representation; these three paths correspond to trails in the dual graph representation (see Definition~\ref{def8}-iii of Appendix~\ref{AP0}). We first note that $X_2 \cup X_3$ forms an Eulerian cycle $G_1$ in the dual graph representation (see Definition~\ref{def8}-v, and Lemma~\ref{eulerian1}), beginning and ending at $S_2$, having $S_1$ as one of its vertices. Because an Eulerian cycle is the union of simple cycles (\cite {harary}, pg. 64) (Figure~\ref{fig:regular-pk-1}-(b)), then the articulation points of $G_1$ have maximum possible degree $4$ (see Corollary~\ref{properties}-2); when we add then the trail $X_1$, from $S_1$  to $S_2$ to $G_1$, $X_1$ cannot touch (include) any of the articulation points of $G_1$. Let $G_2=B_1.a.B_2.b.B_3.c.B_4 \ldots B_r$ (see Definition~\ref{def8}-i, Appendix~\ref{AP0}) be a subgraph of $G_1$ describing a sequence of blocks $B_1, B_2,\ldots, B_r$, $S_1$  is a vertex of $B_1$, and $S_2$ is a vertex of  $B_r$, in which the set $\tilde A=\{a,b,c.....\}$ is the set of articulation points connecting the blocks of $G_2$. Let $G^*$ be the graph obtained by adding the trail $X_1$ to $G_2$. Clearly $\kappa (G^*)$ (see Definition~\ref{def5}-iii) is at least $2$ as deleting a single articulation point in $\tilde A$ won't disconnect $G^*$ as $X_1$ does not have a vertex in $\tilde A$, thus $G^*$ is a non-separable graph (Definition~\ref{def5}-iv). As  both $S_1$ and $S_2$ have degree at least $3$ in $G^*$, then there is a block containing $G^*$ (possibly itself) having a vertex of degree $3$ or more. $\blacksquare$

\subsection {Degrees in Pseudoknots}
\label{AP2}

In this section we introduced the concept of {\it degree} in pseudoknots as defined by Dost et al.~\cite{Dost2008}, to suggest the study of structural properties of RNA secondary structures and of pseudoknots, represented as dual graphs.

\begin{figure}[bth]
\begin{center}
\includegraphics [scale=0.37]{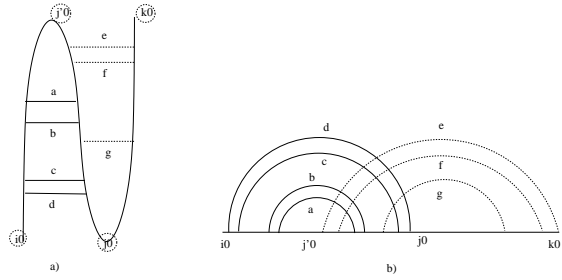}
\end{center}
\caption{Pseudoknot - (a) degree representation of a standard pseudoknot - (b) graphical representation of the pseudoknot where  the $x$-axis is labeled by the primary sequence.}\label{fig:pseudok}
\end{figure}

Intuitively speaking the degree of a pseudoknot refers to the minimum number of {\it turns} (plus one), the primary sequence must be object to avoid the crossing of base pairs. As an example consider Figure~\ref{fig:pseudok}-a, in which two turns (i.e., the pseudoknot is of degree 3) of the primary sequence suffice so no two base pairs intertwine.

Let $ x_1, x_2, \ldots, x_n$ be a sequence of linearly ordered bases.  An alternative notation for a secondary structure is: $ M =\{(i, j) | 1 \leq i < j \leq n, (x_i, x_j)$ is a base pair$\}$. Let also define $ M_{i_0,k_0} \subseteq M$, as $  M_{i_0,k_0}=\{(i, j) \in M | i_0 \leq i < j \leq k_0\}$. The secondary structure, in the absence of crossing or intertwining base pairs is called {\it regular}, and has the following definition. 

\begin{definition} \label{def2} An RNA 2D structure $M_{ i_0 ,k_0}$ is regular iff $M_{i_0 ,k_0} = \emptyset$
or $\exists (i, j) \in  M_{ i_0 ,k_0}$ such that \\
\begin{enumerate}
\item [i.] $ M_{ i_0 ,k_0} = M_{i_0 ,i-1} \cup M_{ i+1,j-1} \cup M_{j+1,k_0} \cup (i, j)$ (no base pairs cross the partitions). 
\item [ii.] Each of $ M_{i_0 ,i-1}, M_{ i+1,j-1}, M_{j+1,k_0}$  is regular. 
\end{enumerate}
\end{definition}

The following two definitions describe the concept of {\it degrees} in pseudoknots (see Figure~\ref{fig:pseudok}). 

\begin {definition}\label{def3} $M_{i_0 ,k_0}$ is a {\it simple pseudoknot}  iff  $M_{i_0 ,k_0}$ is regular or $\exists~j_1 , j_2 \in \Im^+, (i_0 \leq j_1 < j_2 \leq k_0 )$  such that the resulting partition, 
$D_1 = [i_0 , j_1-1], D_2 = [j_1, j_2 - 1], D_3 = [j_2 , k_0 ]$, satisfies the following:\\
\begin{enumerate}
\item[i.] $M_{i_0 ,k_0}= (S_L \cup S_R )$, where $S_L = \{(i, j) \in M_{i_0 ,k_0}| i \in  D_1 , j \in D_2\} $ and $S_R = \{(i, j) \in M_{i_0 ,k_0} | i \in D_2 , j \in D_3 \}$ (i.e., base-pairs crossing regular regions).
\item [ii.] $S_L$ and $S_R$ are regular. 
\end{enumerate}
\end{definition}
\begin {definition} \label{def4} $M_{i_0 ,k_0}$ is a standard ­pseudoknot with degree $d$ ($d \ge 3$) 
iff $M_{i_0 ,k_0}$ is regular or $ \exists~j_1, \ldots, j_d  \in  \Im^+, (i_0 \leq  j_1 < \ldots < j_d \leq k_0 )$ which 
divide $[i_0 , k_0 ]$ into $d$ parts, $D_1 = [i_0 , j_1 - 1], D_2 = [j_1 , j_2 - 1], \ldots, D_d = [j_d , k_0 ]$, 
and satisfy the following: 
\begin{enumerate}
\item [i.] $M_{i_0 ,k_0} =\bigcup_{l=1} ^{d-1} S_l$, where $S_l =\{(i, j)\in M_ {i_0 ,k_0}| i \in D_l , j \in D_{l+1}\},1 \leq l  < d. $
\item [ii.] $S_l$ is regular for all $1 \leq  l < d.$
\end{enumerate}
\end{definition}

Note that a simple ­pseudoknot is a standard ­pseudoknot of degree $3$. 

The concept of degree in pseudoknots can be also extended to the more complex recursive PKs, composed themselves of smaller pseudoknotted regions~\cite{Dost2008,wong}.

\subsection {C++ algorithmic tests performed on RNA motifs mentioned in Section~\ref{S5}}
\label{AP3}
 --------------------- Motif :PDB01069 ----------------------------- \\
===================== New Block ================ \\
(7,5) - (7,4) - (6,7) - (5,6) - (4,5) - (4,5) - \\
degree of 7 is 3 \\
degree of 4 is 3 \\
degree of 5 is 4  \\
 ---- this block represents a pseudoknot ---- \\
===================== New Block ================ \\
(3,4) - \\
---- this block represents a regular-region ---- \\
===================== New Block ================ \\
(3,1) - (2,3) - (2,3) - (1,2) - \\
degree of 3 is 3 \\
degree of 2 is 3  \\
 ---- this block represents a pseudoknot ---- \\
===================== New Block ================ \\
(0,1) - (0,1) - \\
---- this block represents a regular-region ---- \\

----------- Summary information for Motif :PDB01069 ------------------------------\\
     ----------- Total number of blocks: 4 \\
     ----------- number of PK blocks: 2  \\
     ----------- number of regular blocks : 2 \\
------------------------------------------------------------------------------------- \\
--------------------- Motif :PKB236 ----------------------------- \\
===================== New Block ================\\ 
(6,0) - (5,6) - (5,6) - (4,5) - (4,5) - (3,4) - (3,4) - (2,3) - (2,3) - (1,2) - (1,2) - (0,1) - (0,1) - \\
degree of 6 is 3 \\
degree of 5 is 4  \\
degree of 4 is 4  \\
degree of 3 is 4  \\
degree of 2 is 4  \\
degree of 0 is 3  \\
degree of 1 is 4  \\
---- this block represents a pseudoknot ---- \\

----------- Summary information for Motif :PKB236 ------------------------------ \\
      ----------- Total number of blocks: 1  \\
      ----------- number of PK blocks: 1  \\
      ----------- number of regular blocks : 0 \\
------------------------------------------------------------------------------------- \\

\end{document}